\documentclass[pdflatex,sn-mathphys-num]{sn-jnl}\usepackage{graphicx}%
\usepackage{multirow}%
\usepackage{amsmath,amssymb,amsfonts}%
\usepackage{amsthm}%
\usepackage{mathrsfs}%
\usepackage[title]{appendix}%
\usepackage{xcolor}%
\usepackage{textcomp}%
\usepackage{manyfoot}%
\usepackage{booktabs}%
\usepackage{algorithm}%
\usepackage{algorithmicx}%
\usepackage{algpseudocode}%
\usepackage{listings}%

\theoremstyle{thmstyleone}%
\newtheorem{theorem}{Theorem}%

\theoremstyle{thmstyletwo}%
\newtheorem{remark}{Remark}%

\theoremstyle{thmstylethree}%

\newcommand{\bra}[1]{\ensuremath{\left\langle{#1}\right|}}
\newcommand{\ket}[1]{\ensuremath{\left|{#1}\right\rangle}}
\newcommand{\norm}[1]{\ensuremath{\left\lVert{#1}\right\rVert}}
\newcommand{\abs}[1]{\ensuremath{\left\lvert{#1}\right\rvert}}

\begin{document}

\title{Convergence and efficiency proof of quantum imaginary time evolution for bounded order systems}

\author[1,2]{\fnm{Tobias} \sur{Hartung}}\email{tobias.hartung@desy.de}
\equalcont{These authors contributed equally to this work.}
\author*[3,4]{\fnm{Karl} \sur{Jansen}}\email{karl.jansen@desy.de}
\equalcont{These authors contributed equally to this work.}
\affil[1]{\orgdiv{Computing and Information Systems}, \orgname{Northeastern University - London}, \orgaddress{\street{Devon House, St Katharine Docks}, \city{London}, \postcode{E1W 1LP}, \country{United Kingdom}}}
\affil[2]{\orgdiv{Khoury College of Computer Sciences}, \orgname{Northeastern University}, \orgaddress{\street{\#202, West Village Residence Complex H, 440 Huntington Ave}, \city{Boston}, \postcode{02115}, \state{MA}, \country{USA}}}
\affil[3]{\orgdiv{Computation-Based Science and Technology Research Center}, \orgname{The Cyprus Institute}, \orgaddress{\street{20 Kavafi Street}, \city{Nicosia}, \postcode{2121}, \country{Cyprus}}}
\affil[4]{\orgname{Deutsches Elektronen-Synchrotron DESY}, \orgaddress{\street{Platanenallee 6}, \city{Zeuthen}, \postcode{15738}, \country{Germany}}}

\abstract{
  Many current and near-future applications of quantum computing utilise parametric families of quantum circuits and variational methods that can suffer from obstacles including non-convergence to the global minimum due to local minima, critical slowing down, or exponential resource scaling. Here we show that quantum imaginary time evolution can overcome these obstacles if the underlying physical system satisfies a set of conditions. This includes many relevant applications such as ground state preparation for local theories in physics or chemistry, combinatorial optimisation problems, or quantum machine learning. In particular, we analyse the quantum imaginary time evolution showing convergence guarantees to the global minimum without critical slowing down and providing a priori estimates on the required evolution time which scale linearly in system size and inverse energy gap. Furthermore, a provided complexity analysis shows that quantum imaginary time evolution can be efficiently compiled into a parametric quantum circuit, finding the optimal parameters included, for a large class of physically relevant problems. 
}



\maketitle

\section*{Introduction}
With quantum devices becoming ever more powerful, many of their current and near-future applications have similar computational setups. A common approach is to program the quantum device with a parametric quantum circuit (PQC) and to find the correct parameters of the circuit by optimising some cost function through variational methods. Such cost functions may arise naturally in physics~\cite{Feynman,royalsocreview,posreview,ZollerReview,DiMeglio:2023nsa,PhysRevLett.79.2586,doi:10.1126/science.273.5278.1073,PhysRevA.98.032331,PhysRevD.101.074512,PhysRevD.103.094501,PhysRevD.106.114511,RevModPhys.94.015004} or chemistry and drug design~\cite{doi:10.1126/science.1113479,Kandala-Gambetta,RevModPhys.92.015003,https://doi.org/10.1002/qua.26975,8585034,Flother:2024yph}, since the solution of the computational problem is often the ground state of a quantum mechanical system. In fields like quantum machine learning (QML)~\cite{QML,quantum-boltzmann-machines,PRXQuantum.3.030341,PRXQuantum.4.010328,Sauvage_2024,PRXQuantum.5.020328,reptheoryGQML,PhysRevResearch.7.013148,Schatzki,Abbas:2020qrn,PhysRevLett.130.150601,PRXQuantum.4.020327} or combinatorial optimisation~\cite{10.1145/3620668,doi:10.1126/science.1057726,LUXE,FlightGate,Zahedinejad:2017kvw,10.1145/2482767.2482797,morris2024performantneartermquantumcombinatorial,Sankar,Schwagerl:2024xqd,Chicano_2025}, the modelling process commonly involves mapping the problem in question to a quantum mechanical system in such a way that the solution is found by optimising some specifically designed observable. Such observables are commonly related to the Hamiltonian (i.e., energy) of the constructed system.

Since quantum devices generally cannot efficiently measure arbitrary observables, some efficiency assumptions need to be made. A common assumption is locality. For physical systems, this is usually automatically satisfied as the dynamics are commonly described through second order terms such as two-body or nearest-neighbour interactions. While higher order interactions are possible, e.g., plaquette terms in lattice gauge theories are fourth order as four ``link variables'' contribute to each plaquette, most physical systems have some bounded order. Non-physical systems such as those arising from QML or combinatorial optimisation might not have such a ``natural'' order bound, but constructing the problem in such a way that a bounded order system is to be solved on a quantum device is frequently required as otherwise exponential complexity cannot be avoided.

Once a problem is mapped to an instance accessible via variational quantum computing methods, the main challenges are to choose a suitable PQC and a suitable method of finding the required parameters. Here, various approaches are possible. Many PQCs may be chosen because they are efficient to implement on the hardware and/or enforce required symmetries~\cite{DEA,Sauvage_2024,PRXQuantum.4.010328,PRXQuantum.3.030341,wang2023symmetric,Zheng:2022dmx} with optimisers chosen based on performance tests and regularity assumptions of the cost function. Other notable approaches are based on physical system evolutions, such as the quantum approximate optimisation algorithm~(QAOA) which is based on adiabatic evolution~\cite{QAOA2000,QAOA2014}. While adiabaticity has shown to be very successful~\cite{8939749,PhysRevX.10.021067,PhysRevA.104.012403}, it has limitations. For example, the evolution needs to be sufficiently slow to avoid diabatic excitations~\cite{PhysRevA.110.012611}. This can lead to deep circuits with many parameters and highly oscillatory cost function landscapes that are difficult to optimise due to the high density of local optima~\cite{PhysRevResearch.5.023171} and barren plateaus~\cite{s41467-018-07090-4,s41467-024-49909-3,s41467-024-49910-w,McClean2018BarrenPI,Cerezo:2020mtn,BP2025}. Thus, higher-order counterdiabatic terms are required~\cite{RevModPhys.91.045001,Wurtz2022counterdiabaticity,PhysRevA.105.042415,Guan_2025} which can be difficult to implement on current quantum devices.

An alternative to adiabatic evolution is imaginary time evolution (ITE). Although unphysical, ITE is a powerful tool widely used in quantum and statistical mechanics. In particular, in lattice simulations, ITE is known to converge exponentially with evolution time bounded by the inverse energy gap~\cite{Gattringer:2010zz}. Furthermore, quantum ITE implementations have shown good convergence behaviour as well~\cite{McArdle,PhysRevResearch.3.033083,iHVA,symiHVA,IsingQITE,Liu_2022,quantum-boltzmann-machines,Yuan2019theoryofvariational,Motta,PRXQuantum.2.010317,PhysRevResearch.4.033121,Nishi2020ImplementationOQ,PhysRevResearch.6.013143,PRXQuantum.2.010342,PhysRevA.109.052414,PhysRevA.111.012424,osti_1828426,doi:10.1021/acsomega.3c01060,chai2025optimizingquboquantumcomputer}. These are particularly notable for systems with limited interaction terms, e.g., nearest neighbour interactions. While physical intuition suggests a wide applicability range, finding exact conditions under which ITE is efficiently implementable is an open problem. Additionally, the scaling of quantum ITE, that is the implementation of ITE on a quantum device, and whether or not it can circumvent the bottlenecks of other quantum computing approaches remained unknown.

It should be noted that non-variational methods such as quantum phase estimation~\cite{kitaev1995quantummeasurementsabelianstabilizer,PRXQuantum.3.010318,Lin2020nearoptimalground}, exact diagonalisation~\cite{doi:10.1126/sciadv.adu9991,doi:10.1137/21M145954X}, or Krylov space methods~\cite{PhysRevA.105.022417,Kirby2023,Yoshioka2025,yu2025quantumcentricalgorithmsamplebasedkrylov,lee2025generativekrylovsubspacerepresentations} exist as well. These methods have shown remarkable results such as theoretical precision guarantees and exponential convergence with polynomial resources. However, they are also not without challenges. For example, general quantum subspace diagonalisation methods commonly lead to ill-conditioned generalised eigenvalue problems that require suitable truncations. Importantly, for many of these methods provable convergence guarantees or provable polynomial cost seem to require an initial reference state with inverse polynomial fidelity. Since the problem of finding the ground state of local Hamiltonians is QMA-complete~\cite{QMAcomplete}, such restrictions will generally be necessary for any approach and other alternatives to a non-trivial fidelity requirement, including the construction of an adiabatic path from an efficiently solvable Hamiltonian~\cite{QAOA2000}, have been studied as well. In that sense, the choice of computational approach not only depends on the performance of various methods given an instance of a computational problem, but also on relevant a priori and ab initio information.

In this article, we will address some of the above mentioned open questions surrounding convergence guarantees, convergence rate, and compilability of quantum ITE. We will show that quantum ITE retains the exponential convergence observed in lattice theories and that this leads to bounds on the evolution time which are linear in the inverse energy gap. Furthermore, we will show that the evolution time is linear in the number of qubits -- even for exponentially small initial fidelities -- and that no critical slowing down can occur during the evolution. These results are independent of the quantum circuit compilation process which may not be efficiently possible. However, if the Hamiltonian has bounded order and its coefficients have polynomial growth scaling, then we will show that the ITE can be compiled into a PQC of polynomial depth (in the number of qubits and the inverse energy gap), and that the parameters can be found with polynomial cost. Thus, we will prove the efficiency of quantum computations for a large class of bounded order systems which includes many physical and chemical systems, as well as QML and combinatorial optimisation problems.

More precisely, we will prove our main results as summarised in \autoref{thm:main}.
\begin{theorem}\label{thm:main}
  The imaginary time evolution converges with probability $1$ to a state in the lowest energy eigenspace up to bounded error in linear time with respect to the number of qubits and inverse energy gap.
  
  Furthermore, the imaginary time evolution for bounded order systems (defined in Complexity Analysis section) can be compiled into a quantum circuit of polynomial depth in the number of qubits and the compilation cost scales polynomially in the number of qubits and inverse energy gap.
\end{theorem}
The ingredients for the proof of \autoref{thm:main} are split into two categories: General Convergence Results and Complexity Analysis. For the convergence results, we will not consider any means of implementing the ITE, but instead assume that it is possible. We will only discuss circuit compilation in the Complexity Analysis section. Finally, we will provide an explicit Single-Qubit Example highlighting the compilation procedure.

\section*{Results}
\subsection*{General Convergence Results}
For the entire results section, we will assume that the Hamiltonian $H$ is given on a $Q$-qubit system and we choose to order its eigenvalues $\lambda_0\le\lambda_1\le\ldots\le\lambda_N$ increasingly including multiplicities, i.e., $N=2^Q-1$. Denoting the multiplicity of $\lambda_0$ with $\mu$, this means that $\lambda_0=\ldots=\lambda_{\mu-1}<\lambda_\mu$ and we define the energy gap as $\Delta=\lambda_\mu-\lambda_0$. The eigenstates corresponding to the $\lambda_j$ will be denoted $\ket{\psi_j}$. 

\subsubsection*{Convergence Guarantees}
The ITE of given initial state $\ket{\psi(0)}$ with respect to the Hamiltonian $H$ is $\ket{\psi(t)}=\norm{e^{-tH}\ket{\psi(0)}}_2^{-1}e^{-tH}\ket{\psi(0)}$. Using the eigenbasis expansion $\ket{\psi(t)}=\sum_{j=0}^N\alpha_j(t)\ket{\psi_j}$, we observe $\alpha_j(t)=\alpha_j(0)\left(\sum_{k=0}^N|\alpha_k(0)|^2e^{-2t(\lambda_k-\lambda_j)}\right)^{-\frac12}$. This immediately implies $\alpha_j(t)\ne0$ if and only if $\alpha_j(0)\ne0$ for all $t>0$, as well as monotone convergence (in absolute value)
\begin{align}\label{eq:coeffs-conv}
  \alpha_j(t)\to
  \begin{cases}
    \frac{\alpha_j(0)}{\sqrt{\sum_{k=0}^{\mu-1}|\alpha_k(0)|^2}}&,\ j<\mu,\\
    0&,\ j\ge\mu.
  \end{cases}
\end{align}
\autoref{eq:coeffs-conv} assumes at $\alpha_j(0)\ne0$ for at least one $j<\mu$. If that is not the case, the same expression holds for the lowest energy eigenspace that has non-zero overlap with $\ket{\psi(0)}$. In other words, the non-linear operator $\ket{\psi(0)}\mapsto\lim_{t\to\infty}\ket{\psi(t)}$ is equivalent to the two-step process of orthogonally projecting onto the eigenspace of $H$ with smallest eigenvalue that $\ket{\psi(0)}$ isn't orthogonal to and then normalising to the unit sphere in the underlying Hilbert space.

At first sight, this seems to imply that the initial state $\ket{\psi(0)}$ having non-zero overlap with at least one of the states $\ket{\psi_0},\ldots,\ket{\psi_{\mu-1}}$ somehow needs to be satisfied in practice in order to ensure convergence to a state in the lowest energy eigenspace. However, the manifold of states orthogonal to the lowest energy eigenspace is a null set. This follows since the set of states orthogonal to the lowest energy eigenspace is the intersection of the unit sphere of the Hilbert space with a subspace of codimension $\mu$. Thus, unless the Hamiltonian is a multiple of the identity, this orthogonal set is a lower-dimensional submanifold of the state space and Sard's theorem~\cite{sard} implies that all lower-dimensional submanifolds are null sets. The probability for a random initial state to have non-zero overlap with the lowest energy eigenspace is therefore $1$. Furthermore, even if we had such a bad initial state, in the presence of device noise, the probability of remaining in this null set is~$0$. In other words, the ITE converges to a state in the lowest energy eigenspace with probability 1 (up to device precision). Hence, from this point onwards, we will assume non-orthogonality of the initial state $\ket{\psi(0)}$ with the lowest energy eigenspace spanned by $\ket{\psi_0},\ldots,\ket{\psi_{\mu-1}}$.

\subsubsection*{Absence of Critical Slow Down}
Many optimisation algorithms can suffer from a variation of ``critical slow down'', that is, the optimisation appears arrested without having achieved convergence. Considering the path of the evolution $t\mapsto\ket{\psi(t)}$ in the unit sphere $\partial B(0,1)$ of the underlying Hilbert space, we observe
\begin{align}
  \frac{d}{dt}\ket{\psi(t)} = \bra{\psi(t)}H\ket{\psi(t)}\ket{\psi(t)}-H\ket{\psi(t)}=-\mathrm{pr}_{T_{\ket{\psi(t)}}\partial B(0,1)}H\ket{\psi(t)}
\end{align}
where $\mathrm{pr}_{T_{\ket{\psi(t)}}\partial B(0,1)}$ denotes the orthogonal projector onto the tangent space of the unit sphere $\partial B(0,1)$ at the point $\ket{\psi(t)}$. Considering the energy $E(t)=\bra{\psi(t)}H\ket{\psi(t)}$ along the path, this implies $E'(t)=-2\norm{\mathrm{pr}_{T_{\ket{\psi(t)}}\partial B(0,1)}H\ket{\psi(t)}}^2_2\le0$, i.e., the energy is monotonically decreasing. Additionally, the energy gradient can only vanish if $\ket{\psi(t)}$ is an eigenstate of $H$. In other words, any critical slowing down of the evolution can only occur in a neighbourhood of an eigenstate of the Hamiltonian. But, this cannot happen for any excited state $\ket{\psi_j}$ with $\lambda_j>\lambda_0$ due to monotonicity of $\alpha_j\to0$ and, since the energy is efficiently measurable for bounded order Hamiltonians, it provides useful on-the-fly data that can be used to ascertain convergence.

\subsubsection*{Rate of Convergence}
Having established convergence without critical slowing down, i.e., guaranteed convergence in practical settings, we can turn our attention to the rate of convergence. Setting $a=\sum_{k=0}^{\mu-1}|\alpha_k(0)|^2$ and $x=\sum_{k=\mu}^{N-1}|\alpha_k(0)|^2e^{-2t(\lambda_k-\lambda_0)}$ we obtain $\ket{\psi(t)}=\frac{\sum_{k=0}^{\mu-1}\alpha_k(0)\ket{\psi_k}+\sum_{k=\mu}^{N-1}\alpha_k(0)|^2e^{-t(\lambda_k-\lambda_0)}\ket{\psi_k}}{\sqrt{a+x}}$ and $\lim_{\tau\to\infty}\ket{\psi(\tau)}=\frac{\sum_{k=0}^{\mu-1}\alpha_k(0)\ket{\psi_k}}{\sqrt{a}}$. Thus,
\begin{align}\label{eq:rateofconvergence}
  \begin{aligned}
    \norm{\ket{\psi(t)}-\lim_{\tau\to\infty}\ket{\psi(\tau)}}_2^2=&\left(\frac{1}{\sqrt{a+x}}-\frac{1}{\sqrt a}\right)^2a+\frac{x}{a+x}\\
    =&2\left(1-\sqrt{\frac{a}{a+x}}\right)\le\frac{x}{a}\le\frac{1-a}{a}e^{-2t\Delta}.
  \end{aligned}
\end{align}
Thus, we observe exponential convergence with the rate of convergence determined by the energy gap and the constant determined by the initial overlap with the solution.

More precisely, we can bound the fidelity $f(t)=\sum_{j=0}^{\mu-1}|\alpha_j(t)|^2\ge\frac{1}{1+f(0)^{-1}e^{-2t\Delta}}$ which implies that any fidelity threshold $f$ is guaranteed for $t\ge \frac{\ln f-\ln(1-f)-\ln(f(0))}{2\Delta}$. Assuming $f(0)\propto 2^{-Q}$ can be achieved for random or specifically chosen initial states (in the following section we will show that this can reasonably be considered a default assumption for random initial states), this implies that arbitrary fidelity thresholds can be guaranteed in linear time with respect to the number of qubits and inverse energy gap $Q/\Delta$.

\subsubsection*{Bounding the Initial Fidelity}
In the paragraph above, the assumption $f(0)\propto2^{-Q}$ was used to justify linear scaling in $Q/\Delta$ to reach arbitrary fidelity thresholds. Here, we will show that this is generally satisfied and expected for uniformly random initial states in the unit sphere of the underlying Hilbert space, i.e., the quantum device state space.

Let us assume that the lowest eigenvalue has multiplicity $\mu$ and that the initial state is decomposed as $\ket{\psi}=\sum_{j=0}^{2^Q-1}\alpha_j\ket{j}$ where $\{\ket{j};\ 0\le j\le2^Q-1\}$ is an orthonormal basis and $\{\ket{j};\ 0\le j\le \mu-1\}$ an orthonormal basis of the ground state eigenspace. Since a uniform distribution on a sphere in $\mathbb{R}^d$ can be generated by using $d$ independent standard normal distributions to sample the coefficients and normalising~\cite{Muller1,Muller2}, we know that $\ket{\psi}$ can be generated using $2^{Q+1}$ many standard normally distributed random variables $Z_{k,j}$ with $k\in\{1,2\}$ and $0\le j<2^Q$. In fact, we can express the coefficients of $\ket{\psi}$ as $\alpha_j=\frac{Z_{1,j}+iZ_{2,j}}{|Z|}$ with $|Z|=\sqrt{\sum_{j=0}^{2^Q-1}\sum_{k=1}^2Z_{j,k}^2}$.

Setting $\hat Z^2:=\sum_{j=0}^{\mu-1}\sum_{k=1}^2Z_{j,k}^2$ and $\check Z^2:=\sum_{j=\mu}^{2^Q-1}\sum_{k=1}^2Z_{j,k}^2$, the initial fidelity $f(0)$ is then given by the random variable $\frac{\hat Z^2}{\hat Z^2+\check Z^2}$. We also note that $\hat Z^2$ is $\chi^2(2\mu)$-distributed and $\check Z^2$ is $\chi^2(2^{Q+1}-2\mu)$-distributed. Using that the $\mathrm{Beta}(\alpha,\beta)$ distribution can be expressed as $\frac{X}{X+Y}$ with $X\sim\chi^2(2\alpha)$ and $Y\sim\chi^2(2\beta)$~\cite{19d8fbca-9b35-317c-8710-0f3b8f20ffc9}, we conclude that the initial fidelity $f(0)$ has distribution $\mathrm{Beta}(\mu,2^Q-\mu)$. This implies that the expected initial fidelity is $\mathbb{E}(f(0))=\frac{\mu}{2^Q}\propto 2^{-Q}$.

Utilising that the cumulative distribution function of the $\mathrm{Beta}(\alpha,\beta)$-distribution is the regularised incomplete beta-function $I_x(\alpha,\beta)$, we also obtain that the probability of having an initial fidelity less than $\varepsilon$ satisfies 
\begin{align}
  P(f(0)<\varepsilon)=I_\varepsilon(\mu,2^Q-\mu)=\frac{\varepsilon^\mu \Gamma(2^Q)}{\mu\Gamma(\mu)\Gamma(2^Q-\mu)}+O\left(\varepsilon^{\mu+1}\right).
\end{align}
Thus, for $\varepsilon=c2^{-Q}$ with an arbitrary constant $c>0$, we observe
\begin{align}
  P(f(0)<\varepsilon)<\frac{(\varepsilon2^Q)^\mu}{\mu!}+O\left(\varepsilon^{\mu+1}\right)=\frac{c^\mu}{\mu!}+O\left(c^{\mu+1}2^{-Q(\mu+1)}\right).
\end{align}
Hence, we can ensure that the probability $P(f(0)<c2^{-Q})$ is arbitrarily small, i.e., $f(0)\in\Omega(2^{-Q})$ holds with probability $1$. This can be interpreted satisfying $f(0)\propto 2^{-Q}$ or better with probability $1$.

\subsubsection*{Combinatorial Optimisation}
In the case of combinatorial optimisation problems, we can refine the convergence results above. By combinatorial optimisation problems we mean Hamiltonians that are sums of tensor products of Pauli-$Z$ operators. Thus, every individual measurement in the Pauli-$Z$-basis on the quantum device is a viable candidate for the solution. Hence, we generally don't need to ensure that the solution fidelity reaches $1$, but only a threshold $\varepsilon$. E.g., when using $S$ many shots (individual measurements), we can express the threshold fidelity $\varepsilon$ in terms of the number of shots $S$ as $\varepsilon=\kappa/S$ for some $\kappa>0$ which we assume constant when varying $S$. Then, the single-shot success probability at the fidelity threshold is given by $\kappa/S$. This implies that the multi-shot success probability (at least one single-shot success) is given by $1-(1-\kappa/S)^S$ which converges to $1-e^{-\kappa}$ for $S\to\infty$. In other words, a multi-shot success probability $p_S=1-(1-\kappa/S)^S\approx1-e^{-\kappa}$ arbitrarily close to $1$ can be chosen by setting $\kappa=-\ln(1-p_S)$ and choosing $S$ sufficiently large that the error in $(1-\kappa/S)^S\approx e^{-\kappa}$ is negligible.

The initial state can easily be prepared to be an equal superposition of all computational basis states, i.e., $\forall j:\ \alpha_j(0)=2^{-Q/2}$. This implies for the single-shot success probability $p(t)=\sum_{j=0}^{\mu-1}|\alpha_j(t)|^2\ge\frac{1}{1+2^Q\mu^{-1}e^{-2t\Delta}}$. Thus, the multi-shot success probability threshold $p_S=1-e^{-\kappa}$ is approximately reached when $p(t)\ge\varepsilon=\kappa/S$. Finally, this is guaranteed for $t\ge\frac{Q\ln 2-\ln\mu+\ln\varepsilon-\ln(1-\varepsilon)}{2\Delta}$ which scales linearly in $Q/\Delta$.

\subsection*{Complexity Analysis}
Knowing that ITE can reach the solution in linear time up to bounded errors, we still need to consider whether or not the ITE can be efficiently implemented. In full generality, the answer is likely no. However, we will see that it is possible for bounded order systems. Multiple approaches to implementing ITE exist~\cite{Motta,symiHVA,iHVA,McArdle,PhysRevResearch.3.033083,chai2025optimizingquboquantumcomputer}, however for the purposes of this complexity analysis, the finer details are not relevant. Instead, we will use a general circuit construction technique based on dimensional expressivity analysis~\cite{DEA,bestapprox} because it provides additional information that simplifies the complexity analysis.

\subsubsection*{Bounded Order Systems}
We assume a bounded order system, that is, the Hamiltonian is a sum $H=\sum_i a_i S_i$ where each $S_i$ is a bounded tensor product of Pauli-operators and the $a_i$ scale at most polynomially with the number of qubits. This means that there exists a constant $B$ such that all $S_i$ are tensor products of at most $B$ many non-identity Paulis. Hence, there are at most $3^B{Q\choose B}$ many $S_i$ in $H$.

\subsubsection*{Trotterisation}
The time evolution $e^{-tH}$ is then trotterised~\cite{trotter} as $\prod_{\tau=1}^{t/\delta}\prod_i e^{-\delta a_iS_i}$ using small time-steps $\delta$, i.e., there are $O(Q^B\cdot t/\delta)$ many individual factors. The main complexity question therefore becomes whether or not gates acting as the $e^{-\delta a_iS_i}$ can be found efficiently. 

Considering the $e^{-\delta a_iS_i}$ in the $\tau$-layer of the expansion, let $\ket{\psi_{i,\tau}}$ be the corresponding state of the quantum device immediately prior to the $e^{-\delta a_iS_i}$ operation. Up to normalisation, the step of the evolution $e^{-\delta a_iS_i}\ket{\psi_{i,\tau}}$ can then be expressed in terms of a parametric quantum circuit $C_{i,\tau}(\theta_{i,\tau})$ acting on $\ket{\psi_{i,\tau}}$. The exact number of gates in $C_{i,\tau}$ depends on the compilation approach, but a general purpose circuit of depth $5\cdot 2^{B-1}-2$ with $\theta_{i,\tau}\in\mathbb{R}^{2^{B+1}-1}$ can be constructed~\cite{bestapprox} and, using symmetry arguments, much simpler circuits using only up to $4^B$ Pauli-rotation gates (which can be compiled with polynomial depth in the error bound by the  Solovay-Kitaev theorem) but incurring an $O(\delta^2)$-error are known~\cite{iHVA,symiHVA}. These are constant bounds as $B$ is a constant and hence we conclude that the ITE of a Hamiltonian with order bound $B$ can be compiled into a quantum circuit of $O(Q^B\cdot t/\delta)$ depth.

\subsubsection*{Trotterisation Error}
Using the normal Trotter error $e^{-\delta H}-\prod_i e^{-\delta a_iS_i}\in O(\delta^2)$, it is not immediately obvious, how the error propagates when the normalisation is added to each $e^{-\delta a_iS_i}$ step. To analyse this error propagation, we first consider the single-Trotter-step error followed by the integrated error on the interval $[0,t]$. 

For the single-Trotter-step analysis, we first note that 
\begin{align}
  \prod_i C_{i,\tau}(\theta_{i,\tau})\ket{\psi(\tau\delta)}=\frac{\prod_i e^{-\delta a_iS_i}\ket{\psi(\tau\delta)}}{\norm{\prod_i e^{-\delta a_iS_i}\ket{\psi(\tau\delta)}}}=\frac{\prod_i e^{-\delta a_iS_i}\ket{\psi(\tau\delta)}}{\norm{e^{-\delta H}\ket{\psi(\tau\delta)}}+O(\delta^2)}.
\end{align}
Defining $\mathrm{ev}_{\min}(M)$ and $\mathrm{ev}_{\max}(M)$ as the smallest and largest eigenvalues (resp.) of a diagonalisable matrix $M$, we observe $\norm{e^{-\delta H}\ket{\psi(\tau\delta)}}\ge\mathrm{ev}_{\min}(e^{-\delta H})=e^{-\delta \mathrm{ev}_{\max}(H)}$ since $\norm{\ket{\psi(\tau\delta)}}=1$. The polynomial scaling assumption of the $a_i$ in the definition of bounded order systems then implies that $\mathrm{ev}_{\max}(H)\le\sum_i\abs{a_i}\in O(\mathrm{poly}(Q))$. Thus, restricting $\delta$ to $0<\delta\le\hat\delta:=\frac{\ln\gamma}{\sum_i\abs{a_i}}$ for arbitrary $\gamma>1$ yields $\norm{e^{-\delta H}\ket{\psi(\tau\delta)}}\ge\gamma^{-1}$ in a region of no more than polynomially decaying size for $\delta$. The same $\delta\le\hat\delta$ condition also implies $\norm{\prod_i e^{-\delta a_iS_i}\ket{\psi(\tau\delta)}}\ge\gamma^{-1}$ and thus 
\begin{align}
  \begin{aligned}
    \prod_i C_{i,\tau}(\theta_{i,\tau})\ket{\psi(\tau\delta)}=&\frac{\prod_i e^{-\delta a_iS_i}\ket{\psi(\tau\delta)}}{\norm{e^{-\delta H}\ket{\psi(\tau\delta)}}+O(\delta^2)}=\frac{\prod_i e^{-\delta a_iS_i}\ket{\psi(\tau\delta)}}{\norm{e^{-\delta H}\ket{\psi(\tau\delta)}}}+O(\delta^2)\\
    =&\frac{e^{-\delta H}\ket{\psi(\tau\delta)}+O(\delta^2)}{\norm{e^{-\delta H}\ket{\psi(\tau\delta)}}}+O(\delta^2)
  \end{aligned}
\end{align}
which finally implies
\begin{align}
  \norm{\frac{e^{-\delta H}\ket{\psi(\tau\delta)}}{\norm{e^{-\delta H}\ket{\psi(\tau\delta)}}}-\prod_i C_{i,\tau}(\theta_{i,\tau})\ket{\psi(\tau\delta)}}\le\norm{\gamma O(\delta^2)+O(\delta^2)}\le(\gamma+1) O(\delta^2).
\end{align}
In other words, the impact of the normalisation is limited to a change in the constant of the single-Trotter-step error if the step-size has a bound that decays polynomially for bounded order systems.

The integrated Trotter error is then obtained by executing $t\delta^{-1}$ many consecutive single Trotter steps that each have the $(\gamma+1) O(\delta^2)$ bound. Hence, the second Trotter step already starts in a state $\ket{\tilde\psi(\delta)}=\ket{\psi(\delta)}+(\gamma+1) O(\delta^2)$. Considering therefore a $\tau(\gamma+1) O(\delta^2)$ error on the state at time $\tau\delta$, we obtain
\begin{align}
  \norm{\frac{e^{-\delta H}\ket{\psi(\tau\delta)}}{\norm{e^{-\delta H}\ket{\psi(\tau\delta)}}}-\prod_i C_{i,\tau}(\theta_{i,\tau})\left(\ket{\psi(\tau\delta)}+\tau(\gamma+1) O(\delta^2)\right)}\le(\tau+1)(\gamma+1) O(\delta^2).
\end{align}
By induction, the accumulated Trotter error over $\tau$ many steps is $(\gamma+1)\tau O(\delta^2)$, and since there are $\tau=t\delta^{-1}$ many Trotter steps in total, the integrated Trotter error becomes $(\gamma+1) t O(\delta)=O(Q\delta/\Delta)$.

In other words, controlling the Trotter error implies a step-size decay in $O\left((Q/\Delta)^{-1}\right)$ and $O(\hat\delta)=O\left(\frac{1}{\mathrm{poly}(Q)}\right)$, which are both polynomial in $Q/\Delta$.

\subsubsection*{Finding Optimal Parameters}

The circuit compilation requires us to find the optimal values $\theta_{i,\tau}$. For example, this can be done by maximising $\Omega(\delta,\theta_{i,\tau})=\bra{\psi_{i,\tau}}(\cosh(\delta a_i)-\sinh(\delta a_i)S_i)C_{i,\tau}(\theta_{i,\tau})\ket{\psi_{i,\tau}}$. Using the circuit $C_{i,\tau}$ obtained from dimensional expressivity analysis~\cite{DEA} applied to the simpler rotation-gate only circuits~\cite{iHVA,symiHVA}, $\Omega(\delta,\theta_{i,\tau})$ is a polynomial in the variables $s_j=\sin\frac{\theta_{i,\tau,j}}{2}$ and $c_j=\cos\frac{\theta_{i,\tau,j}}{2}$. Thus, $\partial_{\theta_{i,\tau}}\Omega(\delta,\theta_{i,\tau})$ expressed in the $s_j$ and $c_j$ variables together with the conditions $s_j^2+c_j^2-1=0$ defines a system $h(\delta,(s,c))=0$ of polynomial equations. Here, $h$ is a homotopy with trivial solution $h(0,(0,1))=0$ and $\partial_{(s,c)}h(0,(0,1))\ne0$ by construction of $C_{i,\tau}$~\cite{DEA,bestapprox}. Following the proof of the implicit function theorem, polynomial bounds on the step-size $\delta$, which only depend on $B$, $C_{i,\tau}$, and at most $4^B{Q\choose B}$ many $\bra{\psi}\sigma\ket{\psi}$, where $\sigma$ is a Pauli-operator acting non-trivially on at most $B$ many qubits, can be obtained. In fact, using seamless switching~\cite{DEA}, a minimum viable step-size $\delta$ can be guaranteed. Furthermore, the single-shot measurement of $\bra{\psi}\sigma\ket{\psi}$ is a random variable with standard deviation $\le1$, and thus probabilistic error bounds on the multi-shot expectation of $\bra{\psi}\sigma\ket{\psi}$ scale like $(\text{number shots})^{-1/2}$. Hence, the quantum resource requirements scale polynomially with respect to error bound requirements and number of qubits. Using homotopy continuation methods~\cite{Sommese2005,homotopycontinuation}, the circuit for $C_{i,\tau}(\theta_{i,\tau})$ can be compiled in polynomial time, and only polynomially many steps of size $\delta$ are required to reach the total required evolution time $t$. As $t$ scales linearly in $Q/\Delta$, this completes the proof of \autoref{thm:main}.

\subsection*{Single-Qubit Example}
To illustrate the compilation process more clearly, we will consider an explicit single-qubit combinatorial optimisation example. In this case, the Hamiltonian is chosen to be $H=aZ$ and the initial state $\ket{\psi(0)}=\alpha_0(0)\ket0+\alpha_1(0)\ket1$. We chose a combinatorial optimisation Hamiltonian because it allows us to use additional symmetries, namely that we can initialise $\ket{\psi(0)}$ with real amplitudes, e.g., with $\alpha_0(0)=\alpha_1(0)=\frac{1}{\sqrt{2}}$, and since $e^{-taZ}$ is a real operator, $\ket{\psi(t)}$ will always have real amplitudes. Thus, the compiled circuit can be constructed to preserve real amplitudes as well.

In this case, the ITE can be computed analytically, and we obtain $e^{-taZ}\ket{\psi(0)}=\frac{e^{-ta}}{\sqrt2}\ket0+\frac{e^{ta}}{\sqrt2}\ket1$. Adding the normalisation yields $\norm{e^{-taZ}\ket{\psi(0)}}=\sqrt{\cosh(2ta)}$, i.e., 
\begin{align}
  \ket{\psi(t)}=\frac{e^{-ta}}{\sqrt{2\cosh(2ta)}}\ket0+\frac{e^{ta}}{\sqrt{2\cosh(2ta)}}\ket1.
\end{align}

\subsubsection*{Circuit Ansatz}
For the cicuit compilation, we can use dimensional expressivity analysis~\cite{DEA,bestapprox} to obtain that $C(\theta)=R_Y(\theta)$ is a minimal and maximally expressive circuit on the real submanifold of the single-qubit state space manifold. This means that any $\ket{\psi((\tau+1)\delta)}$ can be expressed as $C(\theta_{\tau})\ket{\psi(\tau\delta)}$ (maximally expressive), i.e., that the action of the ITE can be implemented by that circuit, and that any parametric quantum circuit with fewer parameters no longer guarantees the implementability of the ITE (minimality). Of course, in this simple example, dimensional expressivity analysis isn't required to obtain this knowledge, but in general, a minimal and maximally expressive circuit needs to be constructed and dimensional expressivity analysis is a useful tool to achieve that.

\subsubsection*{Homotopy Setup}
Using $C(\theta)=R_Y(\theta)$, the overlap function becomes 
\begin{align}
  \begin{aligned}
    \Omega(\delta,\theta_{\tau})=&\bra{\psi_\tau}(\cosh(\delta a)-\sinh(\delta a)Z)R_Y(\theta_\tau)\ket{\psi_\tau}\\
    =&\cosh(\delta a)\cos\frac{\theta_\tau}{2}-\sinh(\delta a)\bra{\psi_\tau}Z\ket{\psi_\tau}\cos\frac{\theta_\tau}{2}\\
    &-\cosh(\delta a)\bra{\psi_\tau}iY\ket{\psi_\tau}\sin\frac{\theta_\tau}{2}+\sinh(\delta a)\bra{\psi_\tau}X\ket{\psi_\tau}\sin\frac{\theta_\tau}{2}.
  \end{aligned}
\end{align}
Noting that $\ket{\psi_\tau}$ having real amplitudes implies $\bra{\psi_\tau}iY\ket{\psi_\tau}=0$, we can define $\omega_1:=\sinh(\delta a)\bra{\psi_\tau}X\ket{\psi_\tau}$ and $\omega_2:=\cosh(\delta a)-\sinh(\delta a)\bra{\psi_\tau}Z\ket{\psi_\tau}$ to simplify notation to $\Omega=\omega_1\sin\frac{\theta_\tau}{2}+\omega_2\cos\frac{\theta_\tau}{2}$. This overlap function needs to be maximised, i.e., $\theta_\tau$ is a solution to $2\partial_{\theta_\tau}\Omega=\omega_1\cos\frac{\theta_\tau}{2}-\omega_2\sin\frac{\theta_\tau}{2}=0$. Since we also know that $\delta=0$ implies $\theta_\tau=0$, we can understand the overlap condition as a homotopy in $(\delta,\theta_\tau)$. Introducing the substitutions $c=\cos\frac{\theta_\tau}{2}$ and $s=\sin\frac{\theta_\tau}{2}$, this yields the homotopy continuation problem
\begin{align}
  \begin{aligned}\label{eq:homotopy}
    h(\delta,(s,c))=&
    \begin{pmatrix}
      \omega_1c-\omega_2s\\
      s^2+c^2-1
    \end{pmatrix}\\
    h(0,(0,1))=&0.
  \end{aligned}
\end{align}

\subsubsection*{General Homotopy Continuation}
It is possible to solve this homotopy continuation problem analytically. We will do this in the next section. However, in general, numerical methods will be required, so in this section, we will discuss how to numerically solve the homotopy continuation problem.

A reasonably simple approach to solving a homotopy problem is to follow the proof of the implicit function theorem. Doing so, we extend the homotopy by defining $\hat h(\delta,(s,c)):=(\delta,h(\delta,(s,c)))^T$ and noting 
\begin{align*}
  \mathrm{det}\hat h'(\delta,(s,c))=
  \mathrm{det}
  \begin{pmatrix}
    1&0&0\\
    \omega_1'c-\omega_2's&-\omega_2&\omega_1\\
    0&2s&2c
  \end{pmatrix}
  =-2c\omega_2-2s\omega_1.
\end{align*}
Thus, $\mathrm{det}\hat h'(0,(0,1))=-2$ implies invertibility of $\hat h$ in a neighbourhood of $(0,(0,1))$. This neighbourhood depends polynomially on $s$ and $c$, and implicitly on $\delta$ via $\omega_1$ and $\omega_2$. However, the $\delta$-dependence in $\omega_1$ and $\omega_2$ is given by $\cosh(\delta a)$ and $\sinh(\delta a)$ terms which are bounded by constants through the $\delta<\hat\delta=\frac{\ln\gamma}{a}$ condition introduced in the Trotterisation Error section. Thus, the invertibility region is of polynomially decaying size.

\begin{remark}
  The general case as discussed in the Finding Optimal Parameters section has the same structure. The only difference is that there are polynomially many $s_j$ and $c_j$ variables, thus the determinant is a higher order polynomial is $s_j$ and $c_j$. However, the determinant remains a polynomial, and thus the invertibility region remains of polynomially decaying size provided that the determinant is non-zero at the starting point. Here, we could simply confirm the starting point having non-vanishing determinant, but in the general case this is a necessary consequence of the minimality assumption of the circuit ansatz because dimensional expressivity analysis implies that the homotopy image must be lower dimensional if any parameter is removed, i.e., none of the rows or columns in $\hat h'(0,(0,1))$ can be linearly dependent on the rest. 
\end{remark}

It follows that the implicit function $f(\delta)$ with $\hat h'(\delta,f(\delta))=0$ exists and it satisfies 
\begin{align}
  \begin{aligned}
    f'(\delta)=&-\left.\left(\partial_{(s,c)}h(\delta,(s,c))\right)^{-1}\partial_{\delta}h(\delta,(s,c))\right|_{(s,c)=f(\delta)}\\
    f(0)=&(0,1).
  \end{aligned}
\end{align}
This is an implicit ordinary differential equation in two dimensions (in general the dimension is twice the dimension of $\theta_\tau$) and can be solved with a variety of methods. The error scaling here clearly depends on the chosen numerical scheme. For general order bounds $B$ as defined in the Bounded Order Systems section, this also directly leads to precision requirements on the expectation values of the at most $4^B{Q\choose B}$ many $\bra{\psi}\sigma\ket{\psi}$ used to set up the general homotopy in the Finding Optimal Parameters section. 

A minimum viable step-size can be obtained from picking a rectangle that is fully contained in the invertibility region. For example, we may restrict $|\theta_\tau|\le\frac\pi4$, i.e., $0\le|s|\le\frac{1}{\sqrt{2}}\le c\le1$, and compute an upper bound on $\norm{f'(\delta)}$ for $0\le \delta\le\hat\delta=\frac{\ln\gamma}{a}$. Doing so will provide a lower bound $\check\delta$ on $\delta$ for $f(\delta)=(s(\delta),c(\delta))^T$ to reach the $\frac{1}{\sqrt{2}}$ boundaries of $0\le|s|\le\frac{1}{\sqrt{2}}\le c\le1$. The minimum between $\check \delta$ and $\hat\delta$ is then the minimum viable step-size. 

\subsubsection*{Analytic Homotopy Continuation}
While the general homotopy continuation as discussed in the previous section is generally applicable, \autoref{eq:homotopy} can be solved more easily for the purposes of this example. Let us first restrict $|\theta_\tau|<\pi$ for each Trotter step. This implies $c>0$, i.e., $c=\sqrt{1-s^2}$. Then, we simply need to solve $\omega_1c-\omega_2s=0$, i.e., $s^2=\frac{\omega_1^2}{\omega_1^2+\omega_2^2}$. Using $\omega_1(\delta=0)=0$, $\omega_2(\delta=0)=1$, $c(\delta=0)=1$, and $s(\delta=0)=0$, we obtain from $\omega_1c-\omega_2s=0$ that $s$ and $\omega_1$ need to have the same sign near $\delta=0$, i.e., $s=\frac{\omega_1}{\sqrt{\omega_1^2+\omega_2^2}}$. We note that $|s|<1$ holds for all $\delta$. This implies that the restriction $|\theta_\tau|<\pi$ is also satisfied for all $\delta$.

For the first Trotter step, we know that $\ket{\psi_0}=\frac{\ket0+\ket 1}{\sqrt{2}}$, i.e., $\omega_1=\sinh(\delta a)$ and $\omega_2=\cosh(\delta a)$. Using $R_Y(\theta_0)=c-isY$, we obtain
\begin{align}
  \begin{aligned}
    \ket{\psi_1}=&R_Y(\theta_0)\ket{\psi_0}=\frac{(\sqrt{1-s^2}-s)\ket0+(\sqrt{1-s^2}+s)\ket 1}{\sqrt{2}}\\
    =&\frac{e^{-\delta a}}{\sqrt{2\cosh(2\delta a)}}\ket0+\frac{e^{\delta a}}{\sqrt{2\cosh(2\delta a)}}\ket1=\ket{\psi(\delta)}.
  \end{aligned}
\end{align}
Thus, since we didn't have to do any Trotter splitting in this simple example, we obtain that a single Trotter step can be performed for arbitrarily large times and solves the ITE exactly.

\section*{Discussion}
First and foremost, it should be noted that many decisions about the quantum circuit compilation are made in this article because they introduce the necessary regularity providing us with polynomial bounds. These bounds are by no means sharp and better performing circuits can be constructed in practice~\cite{iHVA,symiHVA,DEA,bestapprox,Motta,McArdle,PhysRevResearch.3.033083,PhysRevResearch.4.033121,Nishi2020ImplementationOQ,PhysRevResearch.6.013143,PRXQuantum.2.010342,PhysRevA.109.052414,PhysRevA.111.012424,osti_1828426,chai2025optimizingquboquantumcomputer}.

As such, we have shown that bounded order systems can be solved (up to arbitrary bounded error) using ITE with an evolution time that scales linearly in the number of qubits and inverse energy gap $Q/\Delta$. Furthermore, we have shown that this ITE can be compiled into a quantum circuit with depth scaling polynomially in $Q/\Delta$ and the compilation effort scales polynomially in $Q/\Delta$ as well. This explicit polynomial scaling of the circuit construction, circuit depth, and required evolution time in qubits is what we mean by ``efficient''.

This efficiency claim in terms of qubits needs to be understood in context. Similar to the classical complexity classes P and NP, there are quantum complexity analogues BQP (bounded-error quantum polynomial time) and QMA (Quantum Merlin Arthur or ``quantum NP''). Considering only the linear scaling in qubits without taking the inverse energy gap scaling into account, \autoref{thm:main} at first glance seems to indicate that bounded order systems are in BQP. However, this would be quite an extraordinary claim. The class of bounded order systems considered here contains the class of $k$-local Hamiltonians, and the $k$-local Hamiltonian problem is known to be QMA-complete~\cite{QMAcomplete}, even for $k=2$. Thus, one might hastily conclude BQP=QMA, i.e., the quantum analogue of P=NP. This does not follow from \autoref{thm:main}. Instead, we are exploiting that the physically relevant manifold of states only grows polynomially in the number of qubits for local Hamiltonians which allows for an efficient compilation and simulation on a quantum computer~\cite{PhysRevLett.106.170501}.

The efficiency claim of \autoref{thm:main} is more akin to pseudo-polynomial complexity with respect to the inverse energy gap. The inverse energy gap can have superpolynomial growth in the number of qubits for arbitrary sequences of bounded order systems. Similarly, the assumption that the coefficients $a_i$ have no worse than polynomial scaling is required as otherwise we cannot conclude that the ITE step size $\delta$ scales polynomially. We would only obtain the polynomial scaling for each $\delta a_i$. Otherwise, the Hamiltonian could always be rescaled to have an energy gap $\Delta=1$ which would allow us to prove NP$\subseteq$BQP. Although there is no conclusive evidence, this is unlikely to be true~\cite{BBBU} and it does not follow from \autoref{thm:main}.

Nevertheless, true polynomial scaling is obtained if the sequence of systems has ``well-behaved'' energy gap scaling. This class of systems includes suitable discretisations of a gapped continuum theory, truncations of gapped systems, as well as many combinatorial optimisation problems. Thus, although ``bad'' systems exist, the here shown results are applicable to many highly relevant problems such as in physics~\cite{Feynman,royalsocreview,posreview,ZollerReview,DiMeglio:2023nsa,PhysRevLett.79.2586,doi:10.1126/science.273.5278.1073,PhysRevA.98.032331,PhysRevD.101.074512,PhysRevD.103.094501,PhysRevD.106.114511,RevModPhys.94.015004}, chemistry and drug design~\cite{doi:10.1126/science.1113479,Kandala-Gambetta,RevModPhys.92.015003,https://doi.org/10.1002/qua.26975,8585034,Flother:2024yph}, quantum machine learning~\cite{QML,quantum-boltzmann-machines,PRXQuantum.3.030341,PRXQuantum.4.010328,Sauvage_2024,PRXQuantum.5.020328,reptheoryGQML,PhysRevResearch.7.013148,Schatzki,Abbas:2020qrn,PhysRevLett.130.150601,PRXQuantum.4.020327}, and combinatorial optimisation~\cite{10.1145/3620668,doi:10.1126/science.1057726,LUXE,FlightGate,Zahedinejad:2017kvw,10.1145/2482767.2482797,morris2024performantneartermquantumcombinatorial,Sankar,Schwagerl:2024xqd,Chicano_2025}.

\bibliography{imaginarytimeevolution}

\bmhead{Acknowledgements}
The authors would like to thank Yahui Chai and Stefan Kühn for fruitful discussions. The authors would also like to thank the anonymous reviewer who pointed out an improved rate of convergence argument leading to \autoref{eq:rateofconvergence}.

\bmhead{Funding Statements}
T.H. declares no relevant funding. K.J. discloses support for the research of this work from the European Union’s Horizon Europe Framework Programme (HORIZON) under the ERA Chair scheme with grant agreement no. 101087126 and the Ministry of Science, Research and Culture of the State of Brandenburg within the Centre for Quantum Technologies and Applications (CQTA). 
\begin{center}
  \includegraphics[width = 0.08\textwidth]{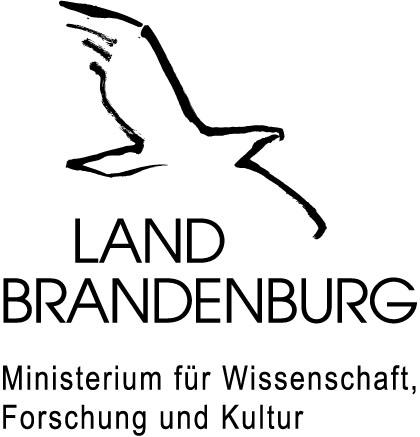}
\end{center}

\bmhead{Competing Interests Statements}
All authors declare no competing interests.

\bmhead{Data Availability Statements}
Data sharing not applicable to this article as no datasets were generated or analysed during the current study.

\bmhead{Author Contributions}
Conceptualization (T.H. lead, K.J. supporting), Methodology (T.H./K.J. equal), Validation (K.J. lead, T.H. supporting), Formal analysis (T.H. lead, K.J. supporting), Investigation (T.H./K.J equal), Writing - Original Draft (T.H. lead, K.J. supporting), Writing - Review \& Editing (T.H./K.J. equal), Funding acquisition (K.J. lead)

All authors have approved the submitted version of the manuscript.

\end{document}